\documentclass[reprint, amsmath, amssymb, aip, jcp, onecolumn]{revtex4-2}

\usepackage{graphicx}
\usepackage{newtxtext}
\usepackage{newtxmath}
\usepackage{natbib}
\usepackage{hyperref}
\hypersetup{
    colorlinks = true,
    urlcolor   = blue,
    citecolor  = black,
}

\newcommand{\RomanNumeralCaps}[1]
\linenumbers

\usepackage[T1]{fontenc}
\usepackage[utf8]{inputenc}

\usepackage[version=3]{mhchem} 

\usepackage{multirow}
\usepackage{dcolumn}
\usepackage{bm}
\usepackage[caption=false]{subfig}
\usepackage[normalem]{ulem} 
\usepackage{makecell} 
\usepackage{booktabs} 
\usepackage{xspace}

\usepackage[per-mode=reciprocal, range-units=single, range-phrase=--]{siunitx} 
\DeclareSIUnit[number-unit-product = {}]\sinumber{{\scriptstyle\#}}

\usepackage{cleveref}
\crefname{figure}{figure}{figures}
\Crefname{figure}{Figure}{Figures}
\crefname{table}{table}{tables}
\Crefname{table}{Table}{Tables}
\crefname{equation}{equation}{equations}
\Crefname{equation}{Equation}{Equations}

\usepackage[textsize=tiny,prependcaption,colorinlistoftodos,textwidth=1.5cm]{todonotes}   

\def \rb {\mathbf{r}}
\def \dr {\, \mathrm{d}\mathbf{r}}

\DeclareSIUnit\bar{bar}
\DeclareSIUnit\angstrom{\text{\AA}}

\def \kb {k_\mathrm{B}}

\makeatletter
\newcommand{\thickhline}{%
    \noalign {\ifnum 0=`}\fi \hrule height 1pt
    \futurelet \reserved@a \@xhline
}
\newcolumntype{"}{@{\hskip\tabcolsep\vrule width 1pt\hskip\tabcolsep}}
\makeatother

\newcommand{\feos}{{FeO$_\mathrm{s}$}\xspace}

\begin{document}

\title{Static Contact Angles of Mixtures: Classical Density Functional Theory and Experimental Investigation 
}

 \author{Benjamin Bursik}
\affiliation{Institute of Thermodynamics and Thermal Process Engineering, University of Stuttgart, Pfaffenwaldring 9, 70569 Stuttgart, Germany}
\author{Nikolaos Karadimitriou}
\affiliation{Institute of Applied Mechanics, University of Stuttgart, Pfaffenwaldring 7, 70569 Stuttgart, Germany}
\author{Holger Steeb}
\affiliation{Institute of Applied Mechanics, University of Stuttgart, Pfaffenwaldring 7, 70569 Stuttgart, Germany}
\author{Joachim Gross}
\affiliation{Institute of Thermodynamics and Thermal Process Engineering, University of Stuttgart, Pfaffenwaldring 9, 70569 Stuttgart, Germany}
 
\begin{abstract}
  This work assesses a classical density functional theory (DFT) model for predicting macroscopic static contact angles of pure substances and mixtures by comparison to own experimental data. 
  We employ a DFT with a Helmholtz energy functional based on the perturbed-chain  statistical associating fluid theory (PC-SAFT) for the fluid-fluid interactions and an effective external potential for the solid-fluid interactions. The solid substrate is characterized by adjusting a single solid-solid interaction energy parameter to a single contact angle value of $n$-octane, while all other results are predictions based on Berthelot-Lorentz combining rules. 
  The surface tensions between solid, liquid and vapor phases are determined from one-dimensional DFT calculations, and Young's equation is used to calculate the contact angle.
  A non-polar polytetrafluoroethylene (PTFE, Teflon) substrate is used, and experiments are carried out using the sessile droplet method with graphical evaluation of the contact angle.
  Accurate results are obtained for pure substances except for monohydric alcohols. 
  Contact angles for monohydric alcohols are systematically overestimated, and we show that this is partially due to neglecting orientational effects. 
The approach provides accurate results for mixtures whenever the respective pure substance contact angle is described well. This comprises mixtures of non-polar, polar, and hydrogen bonding substances, including mixtures with water.
Our results suggest that DFT based on PC-SAFT provides a fast and accurate method for prediction of contact angles for a wide range of pure substances and mixtures.
\end{abstract}

  \maketitle

\section{Introduction}

The wetting behavior of a fluid phase in the presence of a second fluid phase, and brought into contact with a solid, is governed by the interfacial tensions between each pair of the solid, liquid and vapor phase, which result from the interactions between molecules in the fluid and the solid, and possibly external forces, such as gravity. 
Wetting is of high relevance in several technological fields \citep{deGennes2004capillarity,bonn2009wetting,joysula2024fundamentals}. The efficiency of power generation from steam, for example, can be increased by using a hydrophobic coating for the condenser that promotes drop-wise condensation \citep{hoque2023ultra}. Further, the accumulation of water in fuel cells can be prevented by wettability patters which alter the flow field \citep{jang2019enhanced}. Another promising application is the use of superhydrophobic surfaces for harvesting clean water from the atmosphere \citep{miljkovic2013jumping}.

One of the most common methods for characterizing the wetting properties of a system is the measurement of the contact angle at the vapor-liquid and solid-fluid interface. It can vary from $\SI{0}{\degree}$ (total wetting) to $\SI{180}{\degree}$ (total de-wetting) and in between these bounds partial wetting is found. 
Unlike the dynamic contact angle of moving droplets on solid surfaces, which is a function of droplet velocity \citep{narhe2004contact,decker1997contact,koplik1988molecular,lee2022contact,huh1971hydrodynamic,hocking1983spreading,hocking1992rival,eggers2004hydrodynamic}, the static contact angle is an equilibrium property. 
A large variety of approaches exist for determining the static contact angle, both experimental and theoretical. 
Experimental approaches comprise the sessile droplet method, the sliding angle measurements, adhesion force measurements, or the contact angle measurements of fibers (Wilhelmy balance method). The experimental procedure, interpretation and challenges of these methods are discussed in literature \citep{zhao2018contact,hebbar2017contact,lamour2010contact,kwok1999contact,huhtamäki2018surface,decker1999physics,akbari2021contact} and they are applied to various systems \citep{vafei2006effect,sharp2011contact,schuster2015influence,ghasemi2010sessile}.

Theoretical approaches for the calculation of contact angles can be based on macroscopic considerations, such as the Laplace equation \citep{graham2000contact,vafei2005theoretical}, or variational principles \citep{jasper2019generalized}. These approaches typically require input parameters from experiments, or microscopic models. In contrast, molecular simulations provide a predictive framework for determining droplet density profiles and the corresponding contact angle \citep{becker2014contact,hong2009static,nakamura2013dynamic,wang2013contact,lee2022contact,li2018dynamic,yuan2013multiscale,fernandez2019molecular,lukyanov2016dynamic}. They require the solid to be characterized on an atomistic scale, and the quality of predictions depends on a carefully selected force field.

Classical density functional theory (DFT), a method based on statistical mechanics, is also predictive and has low computational cost compared with molecular simulations. It also requires a model for the solid, although molecular roughness, and to some extent also, chemical heterogeneity can be averaged out \citep{eller2021free}.

The equilibrium molecular density is in DFT determined from a variational principle by minimizing  the grand potential functional $\Omega$. It contains the Helmholtz energy functional $F$ for capturing fluid-fluid interactions, the external potential $V^\mathrm{ext}$ for solid-fluid interactions and the imposed chemical potential $\mu$  \citep{evans1979nature,evans2016new}. 
Among the first to study sessile droplets and contact angles with DFT were \citet{talanquer1996nucleation}, who investigated heterogeneous nucleation on a planar solid substrate and determined free energies of formation for critical heterogeneous nuclei and their density profiles. 
In subsequent works, they applied their approach to nucleation and capillary condensation in slit pores \citep{talanquer2001nucleation}, cylindrical capillaries \citep{husowitz2004nucleation} as well as between cylindrical disks \citep{husowitz2005nucleation}. 

A similar approach is based on a lattice version of DFT \citep{malanowski2002lattice,malanowski2002latticeII}.
The approach was applied to analyze experimental results for the interfaces and contact angle of water, on a range of hydrophilic to superhydrophobic surfaces\citep{doshi2005investigating}. Lattice DFT was used to study the influence of surface roughness on the contact angle and droplet nucleation by using pillared surfaces, as well as by comparing results to the macroscopic Cassie and Wenzel equations \citep{porcheron2006mean,guo2014condensation,malanowski2014contact}. Another application of lattice DFT was the calculation of binding potentials\citep{hughes2015liquid}, i.e.\ the contribution of a liquid film to the Helmholtz energy of the system, which determines the wetting state of a given solid-fluid system. A continuum DFT was used to further study binding potentials and the validity of mesoscale models for droplet shape and dynamics that use binding potentials as input \citep{hughes2017influence}. 

Comprehensive studies of sessile droplets using DFT were conducted by \citet{ruckenstein2010microscopic}. 
This approach has been applied to fluids confined in slit pores \citep{berim2008twodimensional,berim2007symmetry} and to studying the relationship between solid–fluid interaction parameters and the contact angle \citep{berim2009simple}. The effects of geometric roughness \citep{berim2011nanodrop,berim2015contact,berim2008nanodrop} and chemical heterogeneity \citep{berim2008nanodrop,berim2009contact} on the contact angle have also been examined. Additional investigations addressed sessile droplets under gravity \citep{berim2008microscopic}, on surfaces with hidden roughness (i.e., roughness covered by a solid overlayer) \citep{berim2015nanodrop}, and on lubricating films \citep{berim2015nanodropLubricating}.

While the previous studies were mostly carried out in two-dimensional coordinate systems, some three-dimensional studies were also conducted. \citet{zhou2012threedimensional} studied heterogeneous nucleation of a Lennard--Jones fluid on solid by using a weighted density approximation in the attractive Helmholtz energy contribution \citep{kim2004density} and, subsequently, investigated line tensions and Tolman lengths\citep{zhou2013line}.  The relation between surface roughness and contact angle, and the validity of the corresponding macroscopic equations was the subject of several works \citep{wang2015evaluation,malijevsky2014does,tretyakov2016cassie,malijevsky2017scaling}. 
A more recent study discusses the conditions for the validity of the Wenzel equation, depending on the shape of surface roughness and the range of solid-fluid interactions by using two-dimensional DFT \citep{egorov2020when}. DFT was also applied to detailed investigations of the three-phase contact region in different geometries \citep{pereira2012equilibrium,nold2014fluid,yatsyshin2015wetting,yatsyshin2018microscopic,nold2018vicinity}, and to bubbles on the microscopic scale \citep{yatsyshin2021surface}. 

The studies discussed so far have significantly advanced the understanding of the factors influencing the contact angle of sessile droplets, and have demonstrated that DFT is a versatile tool for studying such systems. However, most studies are limited to the microscopic scale and/or used model fluids, i.e.\ they did not study contact angles of real substances on the macroscopic scale. 

In contrast, \citet{sauer2018prediction} not only studied microscopic droplets, but also performed a finite-size analysis which showed that extrapolating microscopic contact angles to macroscopic droplet sizes yields good agreement with contact angles predicted by Young's equation.
They extended their analysis beyond model fluids by comparing the contact angles of macroscopic droplets with experimental data for real substances.
This was accomplished by using a Helmholtz energy functional based on the PC-SAFT \citep{gross2001perturbed,gross2002application,gross2002modeling,gross2003modeling,gross2005equation,gross2006equation} model, and  employing a weighted density approximation in the attractive contribution to the Helmholtz energy functional \citep{sauer2017classical}. This approach possesses predictive power for interfacial properties of real substances \citep{gross2009density,rehner2018surface,rehner2021surfactant,nitzke2023phase,bursik2024predicting}, specifically in confined \citep{sauer2019prediction,stierle2024classical,bursik2024viscosities} and even in dynamic systems \citep{stierle2021hydrodynamic,bursik2025modelling}. 
\Citet{sauer2018prediction} only investigated a small number of substances and limited their study to pure substances.  

In this work, we compare DFT predictions for static contact angles of macroscopic droplets to experimental results.  We report contact angles for a larger number of pure substances, with an emphasis on mixtures. The non-polar polytetrafluoroethylene (PTFE, Teflon) is used as the solid substrate.  
The objective of this study is to assess if DFT can provide a fast and accurate method for the prediction of contact angles on the macroscopic scale.

\section{Interfacial Tensions From Classical Density Functional Theory}

We present the essentials of the DFT approach following earlier works \citep{gross2009density,sauer2017classical,sauer2018prediction,sauer2019prediction}. 
DFT provides the molecular density profiles $\rho_i(\rb)$ for each component $i$ in an inhomogeneous system \citep{evans1979nature,evans2016new}.  It is based on the grand potential functional $\Omega$ for specified chemical potentials $\mu_i$, volume $V$ and temperature $T$ as 
 \begin{equation} \label{eq:Omega}
  \Omega\left[\{\rho_i(\rb)\}\right] =F\left[\{\rho_i(\rb)\}\right]                                                 
  -\sum_{i=1}^{N_\mathrm{c}} \int \rho_i(\rb) \left(\mu_i-V_i^\mathrm{ext}(\rb)\right)\dr
\end{equation}
where $N_\mathrm{c}$ is the number of components in the system. The fluid-fluid interactions are modeled by the Helmholtz energy functional $F\left[\{\rho_i(\rb)\}\right]$ with the square brackets and curly brackets denoting a functional dependence and the vector of all components, respectively. The solid-fluid interactions are captured by the external potential $V_i^\mathrm{ext}(\rb)$.

Minimizing the functional in \cref{eq:Omega} leads to 
\begin{equation} \label{eq:ELE}
  \frac{\delta F}{\delta \rho_i(\rb)}-\mu_i+V_i^\mathrm{ext}(\rb)=0 \qquad\qquad \forall i
\end{equation}
This Euler-Lagrange equation allows to  calculate the equilibrium density profiles $\rho_i(\rb)$.

The Helmholtz energy functional used in this work is based on the PC-SAFT model\citep{gross2001perturbed,gross2002application,gross2002modeling,gross2003modeling,gross2005equation,gross2006equation}. It contains additive Helmholtz energy contributions according to 
\begin{equation}
  F  =F^\mathrm{ig}+F^\mathrm{hs} +F^\mathrm{hc} 
  +F^\mathrm{disp}+F^\mathrm{assoc}+F^\mathrm{dp}
\end{equation}
with contributions for the ideal gas $F^\mathrm{ig}$, hard-sphere interactions $F^\mathrm{hs}$, hard-chain formation $F^\mathrm{hc}$, dispersive (non-polar, undirected) interactions $F^\mathrm{disp}$, associative interactions (e.g. hydrogen bonds) $F^\mathrm{assoc}$ as well as dipolar interactions $F^\mathrm{dp}$. The functional dependence $\left[\{\rho_i(\rb)\}\right]$ was not written for brevity.  

The functional for the  ideal gas contribution is known exactly from statistical mechanics. The functionals for the other contributions were derived based on PC-SAFT, which models molecules as chains of tangentially bonded spheres. PC-SAFT requires three pure component parameters for non-polar  components, namely the (real-valued) number of segments per molecule $m_i$, the energy parameter $\varepsilon_{ii}$ and the segment size parameter $\sigma_{ii}$. In addition, associating molecules are described by the association volume parameter $\kappa_{AB}$, and the association energy parameter $\varepsilon_{AB}$. Dipolar molecules are described by the dipole moment $\mu_\mathrm{dp}$.
Modified fundamental measure theory (FMT)\citep{yu2002structures,roth2002fundamental}, which simplifies to the Boubl\'{i}k-Mansoori-Carnahan-Starling-Leland \citep{mansoori1971equilibrium, boublik1970hard} equation of state, is employed for the hard-sphere contribution. For the hard-chain contribution the functional of \citet{tripathi2005microstructure,tripathi2005microstructureII,gross2009density} is used. The functionals for the dispersive and the dipolar contribution were proposed by \citet{sauer2017classical}, based on weighted density approximations. The association contribution is modeled using the functional of \citet{yu2002fundamental}, which is based on Wertheim's thermodynamic perturbation theory of first order (TPT1)\citep{wertheim1984fluids,wertheim1984fluidsII,wertheim1986fluidsIII,wertheim1986fluidsIV}. The pure component parameters for the PC-SAFT model were fitted to experimental vapor-liquid equilibrium data of bulk phases \citep{gross2001perturbed,gross2000application,gross2002application,gross2006equation,esper2023parameters} only, and do not contain interface-specific parameters.

The approach allows for studying mixtures of an arbitrary number of components. The interactions between molecules of different components can be estimated using the Lorentz-Berthelot combining rules, where the interaction energy may be corrected using a binary interaction parameter $k_{ij}$ according to 
\begin{equation} \label{eq:combining}
\begin{aligned}
  \sigma_{ ij}      & = \frac{\left(\sigma_{ii} + \sigma_{jj}\right)}{2} \\
  \varepsilon_{ ij} & = \sqrt{\varepsilon_{ii} \varepsilon_{jj}}(1-k_{ij})
\end{aligned}
\end{equation}
Similarly, the cross-association energy parameter $\varepsilon^{A_iB_j}$ can be determined from a combining rule. However, a recent study \citep{rehner2023modeling} showed that for associating mixtures, an improved description of the binary mixture can be obtained if the   parameter $\varepsilon^{A_iB_j}$ is adjusted to experimental data for phase equilibria of the binary mixtures. This is especially important in the case where induced association occurs: if one component only has a proton acceptor site and no donor site, this leads to a vanishing self-association. However, induced cross-association can still occur with a second component acting as a proton donor. 
 For the binary mixtures studied in this work, the binary interaction $k_{ij}$ and cross-association parameters $\varepsilon^{A_iB_j}$ are taken from \citet{rehner2023modeling}, who adjusted these parameters to vapor-liquid and liquid-liquid equilibrium data for a wide range of mixtures.   

Sessile droplet experiments are performed under atmospheric conditions using a dry nitrogen purge. Thus, nitrogen and possibly traces of oxygen (and other air components) may be present in the system and their influence on the experiments needs to be assessed. One possibility is to perform DFT calculations for a system containing a ternary mixture of nitrogen, oxygen and the substance of interest, as done by \citet{sauer2018prediction}.  However, in the supporting information we show that in DFT calculations the influence of the surrounding atmosphere on the contact angle is negligible. Therefore, we omit the surrounding atmosphere in our calculations, treating the system as either a pure substance or a binary mixture, depending on the context. For pure substances, the temperature is set to the experimental value ($T\approx\SI{298}{\kelvin}$), which fully defines the system, with the resulting pressure equal to the saturation pressure. For binary mixtures, the same temperature is used and the pressure $p=\SI{1}{\bar}$ and the liquid composition are specified.

 PTFE is assumed to be a non-polar substrate, which exhibits only dispersive interactions with the fluids studied here. Defining a representative solid structure of a real PTFE surface with all atomistic positional coordinates is rather difficult. Such positional coordinates are generally unknown and may differ even for substrates from the same material depending on the production process and surface treatment. Instead, we consider the solid as a idealized crystalline solid with a planar interface, where we treat the effective energy parameter as an adjustable parameter. The solid-fluid interactions can then be modeled using an effective, one-dimensional Lennard--Jones 9--3 external potential derived from regular solid-fluid Lennard--Jones pair potentials after integration over a planar solid-fluid interface \citep{sauer2019prediction}. It has proven to be a reliable potential for determining contact angles \citep{sauer2018prediction}. The Lennard--Jones 9--3 potential assumes a uniform solid, and depends only on the distance from the solid, as
 \begin{equation} \label{eq:vext}
  V^\mathrm{ext}_i(z) = \frac{\varepsilon_{si}\sigma^3_{si} \rho_s}{45}\left[ 2\left(\frac{\sigma_{si}}{z}\right)^9  - 15\left(\frac{\sigma_{si}}{z}\right)^3\right]
 \end{equation} 
 where $\varepsilon_{si}$ and $\sigma_{si}$ are the solid-fluid interaction energy and diameter parameter, respectively, for fluid component $i$. $\rho_s$ is the number density of interaction sites in the bulk solid. For the solid-fluid interaction parameters we also assume Lorentz-Berthelot combining rules, \cref{eq:combining}. Thus, knowledge of the parameters of the solid $\varepsilon_{ss}$, $\sigma_{ss}$ and $\rho_s$ is required. As in  \citet{sauer2018prediction}, we define  $\sigma_{ss}= \SI{3.0}{\angstrom}$ and $\rho_s=\SI{0.08}{\per\angstrom\cubed}$ for the diameter parameter and the solid density, respectively. 
 The energy interaction parameter $\varepsilon_\mathrm{ss}$ is adjusted to a single measurement of the contact angle of $n$-octane, since $n$-octane represents a non-polar substance where any solid-fluid interactions due to non-dispersive (i.e.\ polar or associating) effects can be excluded. Alternatively, the parameter could be adjusted to contact angles of several non-polar substances at the same time, which provides similar numerical values for the substances studied here. Besides the pure and binary PC-SAFT parameters for the fluids, which are available for a large number of substances in literature\citep{esper2023parameters,rehner2023modeling}, no adjustable parameters enter the contact angle model. Beyond $n$-octane, we consider the DFT approach to be predictive with regard to contact angles.

\subsection{PC-iSAFT: A Helmholtz Energy Functional Distinguishing Individual Interactions Sites of Fluids}
According to the molecular model of PC-SAFT, molecules are composed of spherical segments connected to chains.
The  DFT approach presented so far considers the densities of all segments in an averaged way, without distinguishing the difference between, say, a terminal segment and a mid-of-molecule segment.
This approach reduces the complexity and computational cost of the model at the expense of a lower level of physical detail. The theory of Wertheim allows for a more detailed modeling approach, where the connectivity of segments is accounted for \citep{wertheim1986fluidsIII,wertheim1986fluidsIV, zmpitas2016detailed}. The iSAFT formalism\citep{jain2007modified,rehner2021surfactant}  provides the densities of individual segments of a molecule. This model thus, allows to capture the orientation of molecules to a certain extent. We refer to the original publication \citep{jain2007modified} and the adaption to the PC-SAFT model \citep{mairhofer2018classical,rehner2021surfactant} for additional information. 

The DFT calculations are performed using the open-source thermodynamic software package \feos \citep{rehner2021application,rehner2023feos}. The one-dimensional grid has a length of $\SI{150}{\angstrom}$ and 2048 grid cells are used, which results in a grid resolution with negligible discretization error. For details on the implementation of DFT we refer to \citet{stierle2020guide,stierle2024classical}.

\subsection{Contact Angles from Young's Equation}

The thickness of the three interfaces (vapor-liquid, solid-liquid and solid-vapor) is very small compared to the size of a  sessile droplet on the macroscopic scale. Consequently, the sharp interface approximation is reasonable and Young's equation can be applied \citep{sauer2018prediction} according to 
\begin{equation}
  \cos \Theta = \frac{\gamma_\mathrm{sv}-\gamma_\mathrm{sl}}{\gamma_\mathrm{vl}}
\end{equation} 
with the static contact angle $\Theta$, as well as the interfacial tensions between vapor-liquid $\gamma_\mathrm{vl}$, solid-liquid $\gamma_\mathrm{sl}$ and solid-vapor $\gamma_\mathrm{sv}$ phases. The interfacial tensions can be determined from three independent one-dimensional DFT calculations. For each interface \cref{eq:ELE} is solved for the density profile and the interfacial tensions  can be determined from
\begin{equation}
   \gamma_{\alpha\beta} = \frac{\Omega_{\alpha\beta}}{A} + L_z p
\end{equation}
where $L_z$ is the length of the considered (planar) system, $p$ is the equilibrium bulk pressure (which is equal to the normal component of the pressure tensor) and the grand potential $\Omega_{\alpha\beta}$ for a system containing the interface between phases $\alpha$ and $\beta$  is evaluated using \cref{eq:Omega}. 
 In each case, it is assumed that the interfacial tension of the macroscopic droplet is equal to the interfacial tension of the equivalent planar system. This effectively introduces the capillarity approximation, which is typically used in classical nucleation theory \citep{kalikmanov2013nucleation}. For microscopic droplets, curvature-dependent surface tensions can be determined based on Tolman lengths using the same Helmholtz energy functional as in this work \citep{rehner2018surface}. 

\section{Contact Angles Experiments}

\begin{figure}
  \centering
  \includegraphics[width=0.75\textwidth]{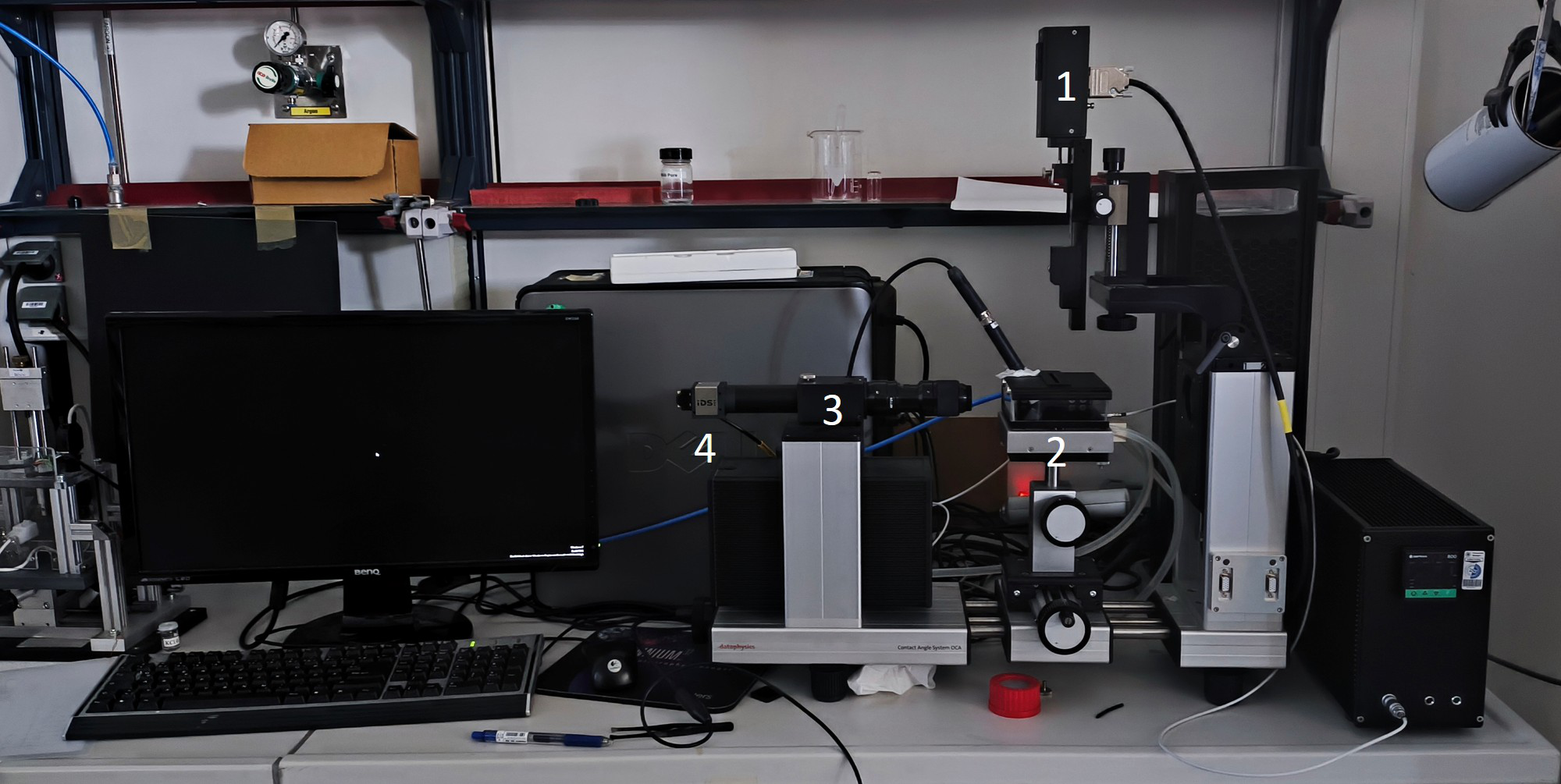} 
  \caption{Picture from the experimental setup, with the major components numbered. 1.) The motorized stage for the disposal of the droplet. 2.) The chamber with controlled humidity and temperature, where the droplet is disposed on the sample. 3.) The focusing objective lens. 4.) The IDS\textcopyright camera.}
  \label{fig:setup}
\end{figure}

\begin{figure}
  \centering
  \includegraphics[width=0.45\textwidth]{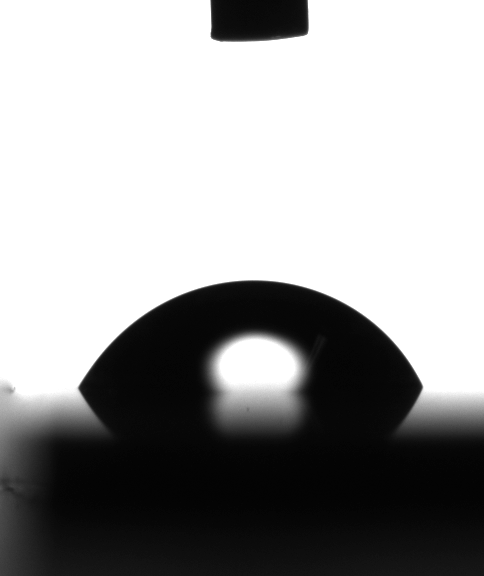} 
  \caption{A picture as it was taken from the setup showing a droplet on the substrate's surface. From such an image the contact angle on both ends of the droplet was calculated using the dedicated software (SCA20).}
  \label{fig:droplet}
\end{figure}

Experiments on sessile droplets on a PTFE substrate are conducted, with three separate series of experiments. 
Measurements were performed using the OCA25 system developed by dataphysics\textcopyright, as shown in  \cref{fig:setup}. This setup allowed for controlled dispensing of liquid volumes through a dedicated software (SCA20), precise sample positioning along all three spatial axes using high-precision mechanical stages, and fixed-focus imaging with adjustable magnification up to 6.5×. The setup also offered the ability to control the humidity and temperature, with the use of nitrogen and a heat plate respectively. Image acquisition was performed using a monochrome IDS\textcopyright UI-2220SE-M camera at a resolution of 768 × 576 pixels. Before each measurement, the substrate was initially cleaned in an ultrasonic bath with acetone for 90 seconds, and then it was blown dry with nitrogen to remove any potential dust from the surface of interest. For all measurements, droplets with a volume of \SI{6}{\micro\liter} were used, as this volume provided optimal visualization with minimal measurement error. The droplets were dispensed through a needle with a diameter of 0.5 mm, at a rate of \SI{1}{\micro\liter\per\second}. An image from the droplet was captured, like the one shown in \cref{fig:droplet}, and by using the same software environment as for the disposal of the volume of the fluid, the contact angle was calculated automatically. All experiments were carried out at atmospheric pressure ($p\approx\SI{1}{\bar}$).

The reproducibility of the experimental contact angles is satisfactory with a standard deviation of about \SI{1.7}{\degree} for measurements of a given liquid (pure or mixture) during an experimental run. However, between different experimental runs (i.e. after the substrate has been exposed to other substances or mixtures), somewhat larger differences in the contact angle were observed for the same substance, with a standard deviation about \SI{3.0}{\degree}. The increase of standard deviation is likely caused by the strong sensitivity of the contact angle towards the specific state of the solid surface, which might change due to contact with the different substances during the experiments, or due to our cleaning and drying procedure. We therefore chose to evaluate the three experimental runs separately, i.e.\ we adjust the solid energy parameter to the contact angle of $n$-octane separately for each experimental run. This is expected to provide the most accurate basis for evaluation of the DFT approach.
In addition to our own experiments, \citet{li1992contact} published contact angles obtained from experiments similar to the ones conducted here and this data was also used for comparing predictions from DFT to experimental data. 

\section{Results and discussion} \label{sec:results}

\subsection{Contact Angles of Pure Substances}

\begin{figure}
  \centering
  \includegraphics[width=0.6\textwidth]{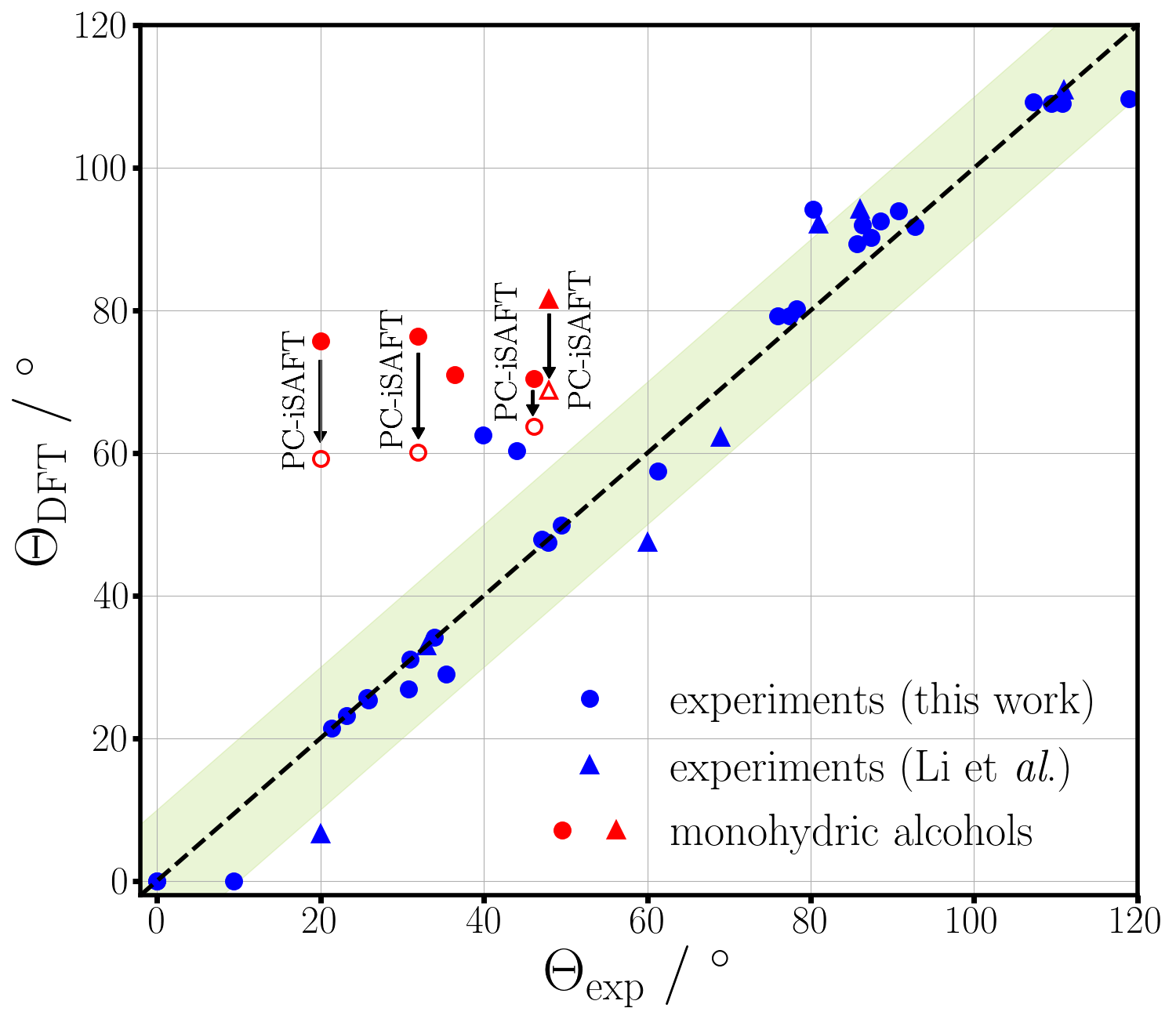} 
  \caption{Comparison of static contact angles from DFT and experiments (this work and from \citet{li1992contact}) for pure substances on solid PTFE substrate at $T\approx\SI{298}{\kelvin}$. Red symbols for monohydric alcohols; blue symbols for all other substances. The green area indicates deviations of less than $10^\circ$. Red open circles show results from PC-iSAFT\citep{jain2007modified,rehner2021surfactant} model for monohydric alcohols. Note that some components appear multiple times in the diagram, since they were measured repeatedly in the different experimental runs. 
  }
  \label{fig:exp_pure}
\end{figure}
\begin{table*}
  \centering
  \caption{Static contact angles from DFT in comparison to experimental data of this work (top part) as well as experiments from \citet{li1992contact}(lower part) for pure substances on solid PTFE substrate at $T\approx\SI{298}{\kelvin}$. }
  \label{table:exp_angles}
  \renewcommand{\arraystretch}{1.2}
  \begin{tabular}{lrr @{\hskip 1cm} l rr}
    \toprule
    Component & $\Theta_\mathrm{Exp} ~/~ ^\circ$ & $\Theta_\mathrm{DFT} ~ / ~ ^\circ$ & Component & $\Theta_\mathrm{Exp} ~/~ ^\circ$ & $\Theta_\mathrm{DFT} ~ / ~ ^\circ$ \\
    \midrule
     \multicolumn{6}{l}{\underline{Own experiments (this work)}} \\
tetralin & 61.3 & 57.5 & DMSO & 85.7 & 89.4 \\
1,2-ethanediol & 86.3 & 92.0 & DMSO & 87.3 & 90.3 \\
1,2-ethanediol & 92.8 & 91.7 & hexadecane & 47.9 & 47.4 \\
1,2-ethanediol & 88.5 & 92.5 & hexane & 0.0 & 0.0 \\
1,2-propanediol & 80.2 & 94.2 & NMP & 77.4 & 79.2 \\
1-pentanol & 46.1 & 70.4 & NMP & 76.0 & 79.2 \\
2,2,4-trimethylpentane & 0.0 & 0.0 & NMP & 78.3 & 80.2 \\
2-butanone & 44.0 & 60.3 & octane & 25.7 & 25.7 \\
2-methyl-1-propanol & 36.4 & 71.0 & octane & 21.4 & 21.4 \\
2-propanol & 31.9 & 76.4 & octane & 23.2 & 23.2 \\
2-propanol & 20.0 & 75.7 & pentane & 9.4 & 0.0 \\
acetone & 39.9 & 62.5 & toluene & 49.5 & 49.9 \\
cyclohexane & 30.8 & 26.9 & toluene & 47.1 & 47.9 \\
cyclohexane & 25.9 & 25.4 & water & 109.4 & 109.0 \\
cyclohexane & 35.4 & 29.0 & water & 110.8 & 109.0 \\
dibutyl ether & 34.0 & 34.2 & water & 118.9 & 109.6 \\
dibutyl ether & 31.0 & 31.1 & water & 107.3 & 109.3 \\
triethanolamine & 90.8 & 94.0 &  \\
    \midrule
    \multicolumn{6}{l}{\underline{Literature data (\citet{li1992contact})}} \\
    tetralin & 69 & 62.2 & DMSO & 81 & 92.1 \\
1,2-ethanediol & 86 & 94.2 & hexane & 20 & 6.6 \\
1-propanol & 48 & 81.6 & octane & 33 & 33.0 \\
decalin & 60 & 47.5 & water & 111 & 110.9 \\
    \bottomrule
    \end{tabular}
\end{table*}
We first consider several pure substances in \cref{fig:exp_pure}, where
experimental results of this work are shown for three experimental runs. Data from \citet{li1992contact} is included, for which $\varepsilon_\mathrm{ss}$ (characterizing the PTFE solid) was adjusted individually to measurements for $n$-octane.
The resulting $\varepsilon_\mathrm{ss}$-values, as determined by our experiments for three independent runs, were $\varepsilon_\mathrm{ss}/\kb  \in \{\SI{69.31}{\kelvin}, \SI{68.376}{\kelvin}, \SI{69.94}{\kelvin} \} $, which suggests a good reproducibility of the experiments. A slightly different value of $\varepsilon_\mathrm{ss}/\kb = \SI{65.19}{\kelvin} $ was determined for the data from \citet{li1992contact}, which might be due to differences in composition and structure of the solid material that was used. 
Numerical values for the pure component contact angles are provided in \cref{table:exp_angles}.

For most substances, predictions from DFT agree favorably with experimental values, with deviations in contact angles below  $10^\circ$. 
However, for monohydric alcohols, i.e.\ alcohols with exactly one hydroxyl group, much larger deviations are observed. Two reasons may contribute to this result: First, the solid PTFE substrate might actually contain some polar sites that can form specific dipolar interactions, or hydrogen bonds with the alcohols, which is in contrast to the assumption of a purely non-polar substrate. Such interactions are not included in the external potential, and would explain the overestimation of contact angles from DFT. A second explanation is that the orientation of these substances at the solid-fluid interface is not properly accounted for, because the DFT approach employed here does not resolve molecular orientation. Monohydric alcohols possess a polar hydroxyl group and a non-polar part, which leads to a preferred orientation of the molecule where the non-polar part is oriented towards the non-polar solid, and the polar part is oriented in the opposite direction.  

The second hypothesis can be further examined by calculating results using the PC-iSAFT model\citep{jain2007modified,rehner2021surfactant}. This approach assumes that the molecules consist of different segments with defined connectivity, for which individual density profiles are evaluated. 
Although this approach does not explicitly account for precise orientational angles of individual molecules, it effectively captures the  orientation of molecular segments at interfaces.
While the PC-iSAFT model reduces the deviations observed for monohydric alcohols (see \cref{fig:exp_pure}), the agreement with experimental data for these substances remains unsatisfactory.

The investigation of pure substances shows that the considered DFT model is capable of predicting contact angles for a wide range of substances, including linear and cyclic alkanes, dihydric alcohols (e.g.\  1,2-ethanediol), and water. The good agreement of the method for the contact angle of water is remarkable and somewhat unexpected, because water is typically difficult to model, and is also subject to specific orientational distributions and possibly specific polar interactions towards the solid. Excluding monohydric alcohols and $n$-octane (which was used in the parameter adjustment), the  mean absolute deviation for all substances is about \SI{5.1}{\degree}. 
For the following investigations of mixtures, we exclude monohydric alcohols and focus on mixtures of the remaining substances. 

\subsection{Contact Angles of Mixtures}

\begin{figure}
  \centering
  \includegraphics[width=0.6\textwidth]{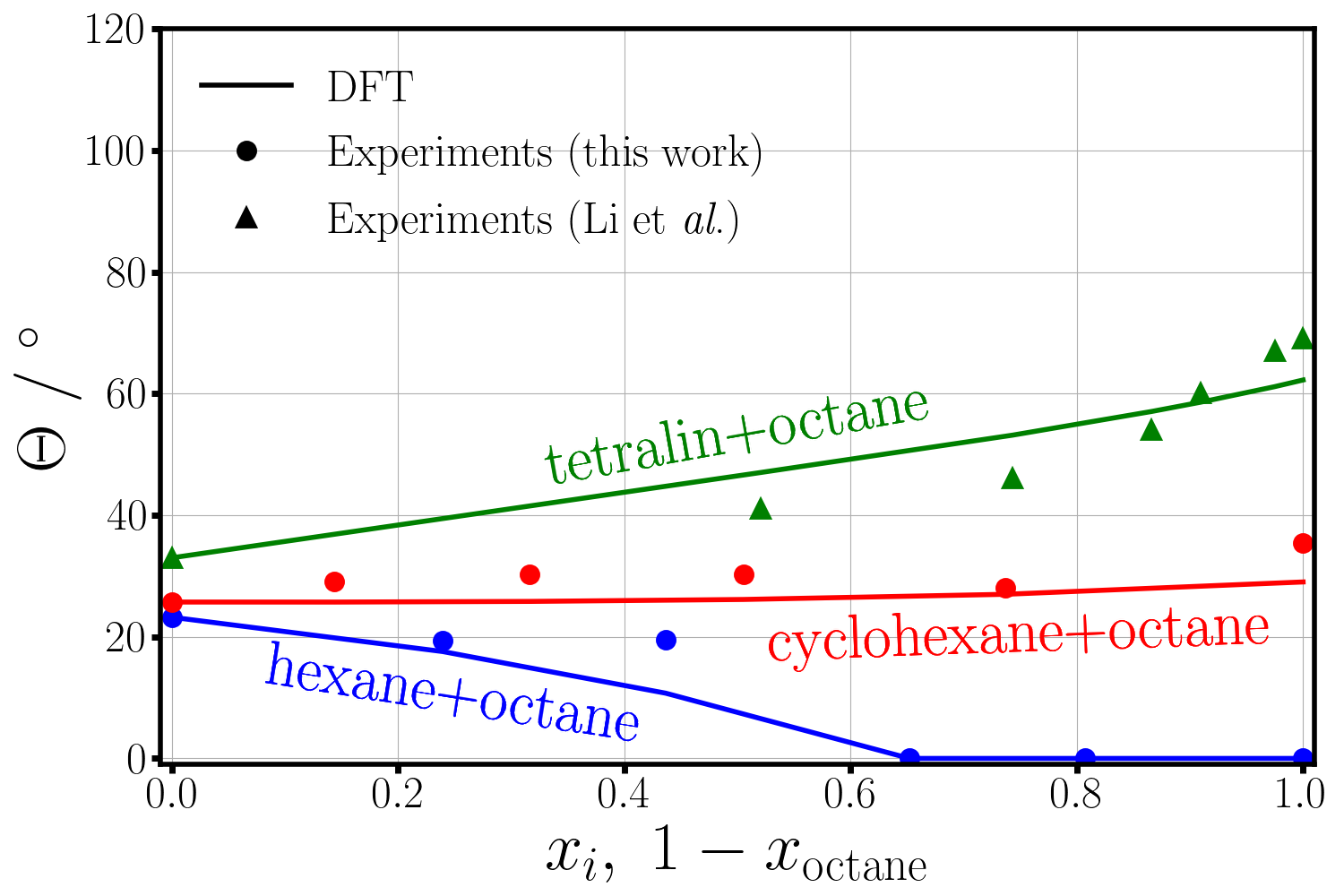}
  \caption{Static contact angles  of binary mixtures of $n$-octane with $n$-hexane, cyclohexane and tetralin, respectively, on solid PTFE substrate for different molar compositions. Predictions from DFT are compared to experimental data (this work and \citet{li1992contact}),  at $T\approx\SI{298}{\kelvin}$. 
  }
  \label{fig:exp_mix_octane}
\end{figure}
In this section, we compare contact angles of mixtures predicted from DFT to experimental data. \Cref{fig:exp_mix_octane} shows contact angles for binary mixtures of different substances and $n$-octane for varying composition. Note that the results from DFT and experiments  agree exactly for pure $n$-octane, because $n$-octane was used for parameter adjustment. Furthermore, experimental values are slightly different because they are obtained from different experimental runs and from literature data \citep{li1992contact}. 
Starting from pure $n$-octane, an increasing mole fraction of $n$-hexane leads to a decreasing contact angle. At approximately $x_\mathrm{hexane}=0.6$ the contact angles become zero, i.e.\ total wetting is obtained. 
For increasing mole fractions of cyclohexane, the contact angle of the mixture increases mildly. For tetralin, the contact angle increases substantially and reaches values above \SI{60}{\degree}. 
In all cases, the DFT approach captures the qualitative trends and is in satisfactory quantitative agreement with experimental results. 

\begin{figure}
  \centering
  \includegraphics[width=0.6\textwidth]{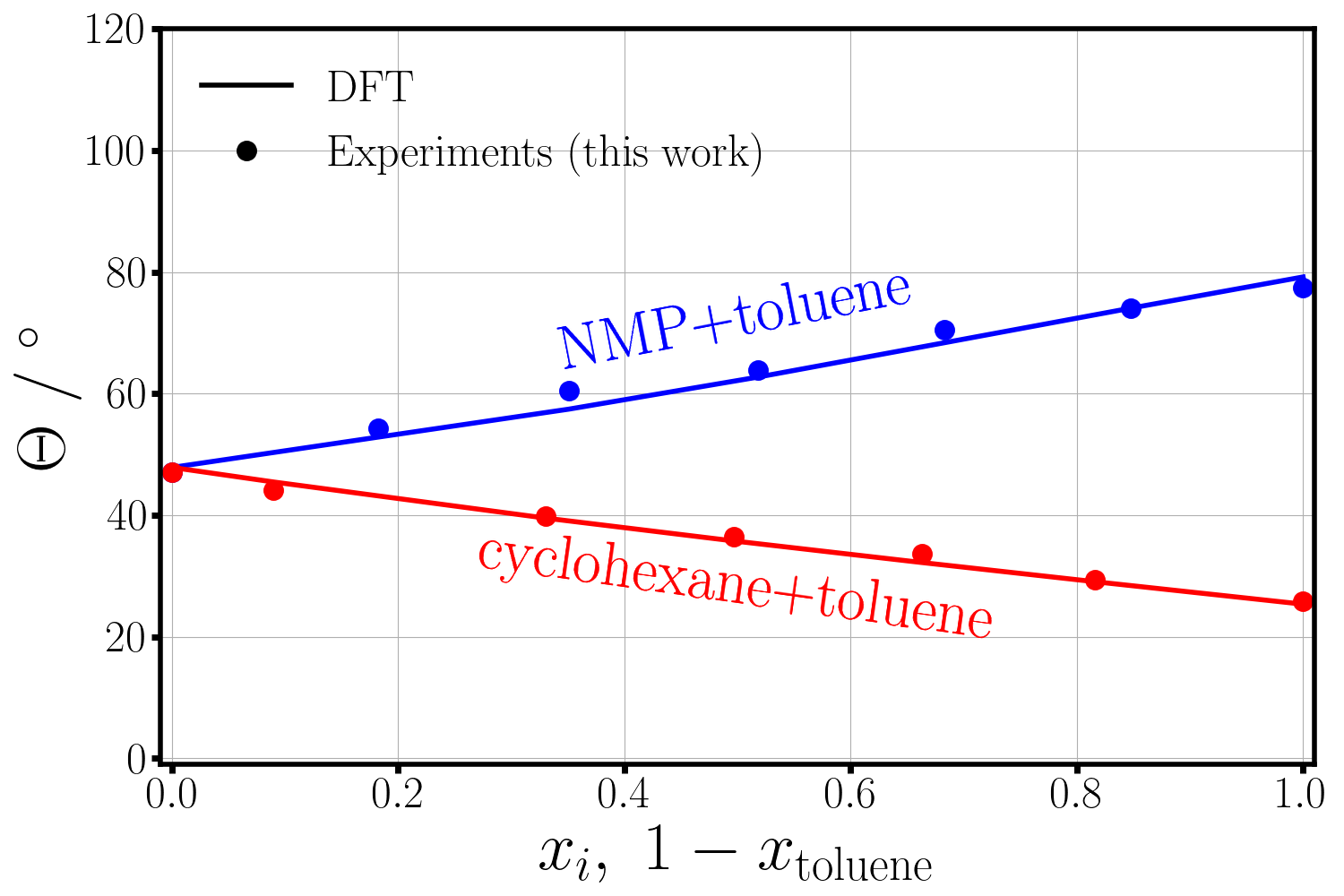}
  \caption{Static contact angles for binary mixtures of toluene with $n$-methyl-2-pyrrolidone (NMP) and with cyclohexane, respectively,  on PTFE for different molar compositions. Predictions from DFT are compared to experimental data, at $T\approx\SI{298}{\kelvin}$. 
  }
  \label{fig:exp_mix_toluene}
\end{figure}
\cref{fig:exp_mix_toluene} summarizes results for mixtures of toluene with NMP and with cyclohexane, respectively. $n$-methyl-2-pyrrolidone (NMP) is an organic aprotic solvent, i.e.\ it is polar but can only act as a proton acceptor, and it is here modeled as a dipolar but non-associating component. Note that none of these substances were used in the adjustment of the solid parameters and, thus, contact angle results from DFT are predictions. The contact angle of pure toluene is larger than this for pure $n$-octane. For an increasing mole fraction of NMP the contact angle increases, while it decreases for cyclohexane. In both cases, predictions from DFT are in excellent agreement with experimental results. 

\begin{figure}
  \centering
  \includegraphics[width=0.6\textwidth]{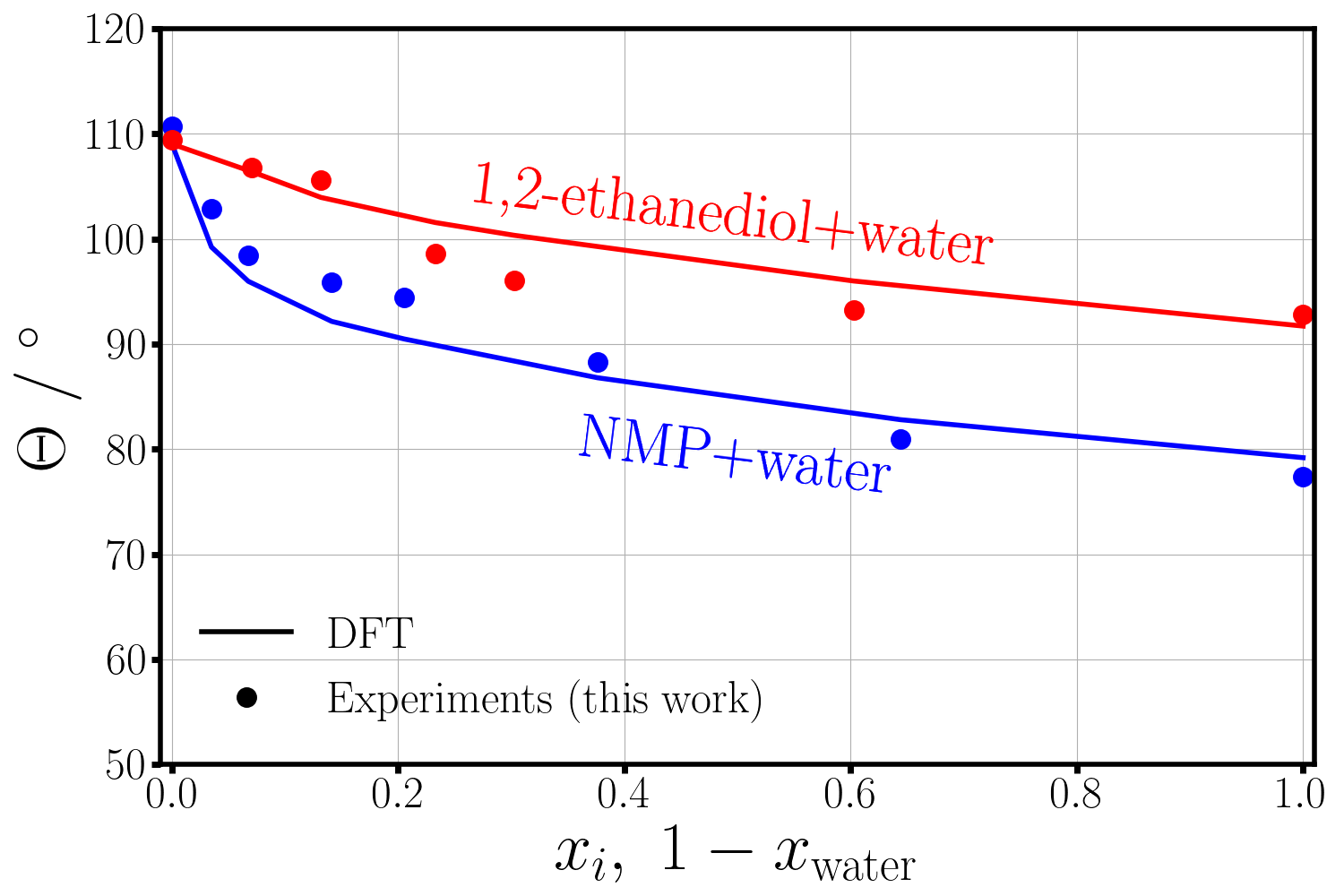}
  \caption{Static contact angles  of binary mixtures of  water with NMP and with 1,2-ethanediol, respectively, on PTFE for different molar compositions. Predictions from DFT are compared to experimental data, at $T\approx\SI{298}{\kelvin}$. 
  }
  \label{fig:exp_mix_water}
\end{figure}

The results for mixtures presented so far, comprise non-polar or polar components which do not exhibit hydrogen bonds (e.g.\ NMP). The pure component and mixture contact angles of these substances in contact with the (mostly) non-polar PTFE are, therefore, below \SI{80}{\degree}. \Cref{fig:exp_mix_water} shows results for mixtures with water, which has a pure component contact angle of about \SI{110}{\degree}. This is the largest contact angle of all substances investigated here. The diagram shows mixtures of water with 1,2-ethanediol and with NMP, respectively. For water with 1,2-ethanediol, both pure substances exhibit self-association (hydrogen bonding), and cross association occurs in the mixture. For water with NMP, NMP does not self-associate but it is capable of cross-associating (accepting protons) with water. This effect is captured by including cross association parameters obtained from \citet{rehner2023modeling}. 

For both mixtures, the contact angle decreases with increasing mole fraction of the second component. Interestingly, for the mixture of water and NMP this decrease in the contact angle appears at relatively low mole fractions of NMP. The molar weight of NMP is approximately 5.5 times larger than the molar weight of water (\SI{99.068}{\gram\per\mole} vs. \SI{18.011}{\gram\per\mole}) and consequently, a small mole fraction of NMP corresponds to a larger mass fraction. This suggests that, in this case, the contact angle would change in a more linear manner when visualized using mass fractions instead of mole fractions. 
The DFT approach captures this effect and is in good quantitative agreement with experimental results. The results indicate that the DFT approach is capable of accurately predicting contact angles for  mixtures of associating (hydrogen bonding) substances. 

\begin{figure}
  \centering
  \includegraphics[width=0.6\textwidth]{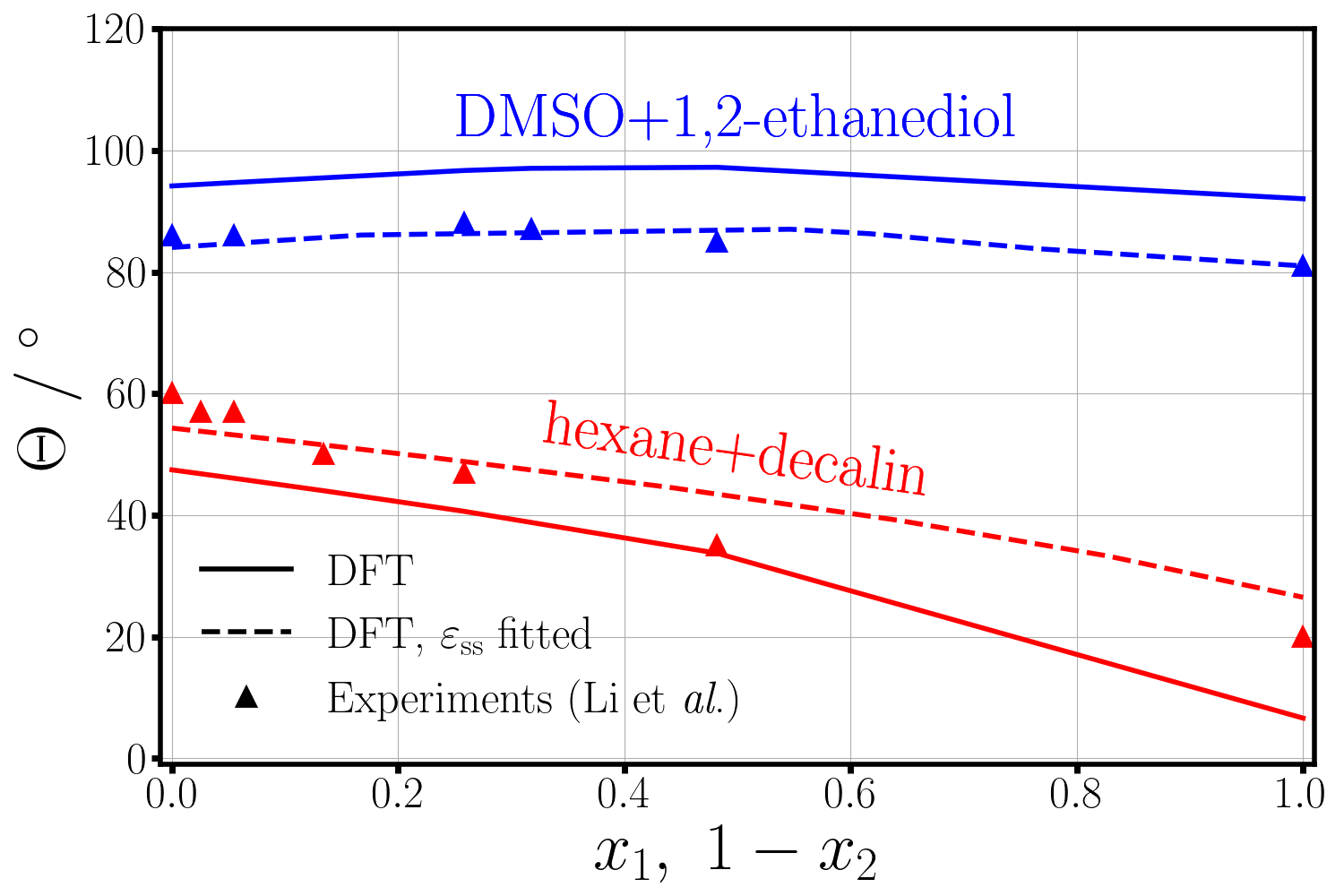}
  \caption{Static contact angles of two binary mixtures on a solid PTFE substrate for varying molar compositions. Results from DFT and experiments from \citet{li1992contact}  at $T\approx\SI{298}{\kelvin}$. 
  }
  \label{fig:exp_mix_rest}
\end{figure}
Finally, results for two mixtures, where the pure component contact angles exhibit deviations close to or above \SI{10}{\degree},  are shown in \cref{fig:exp_mix_rest}. The solid lines are DFT predictions, analogously to the previous figures, and crosses are experimental values.
In order to isolate deviations of the DFT model for mixtures from deviations of the model for pure substances, we include dashed lines for DFT results, where the energy interaction parameter of the solid $\varepsilon_\mathrm{ss}$ was readjusted  to the pure component contact angles of the two constituents of each mixture. For the mixture of $n$-hexane and decalin, even though at $x_\mathrm{hexane}\approx 0.5$ the experimental contact angle deviates somewhat from the DFT results with readjusted $\varepsilon_\mathrm{ss}$, the overall agreement with experiments is improved. For the mixture of dimethyl sulfoxide (DMSO) and 1,2-ethanediol the agreement of DFT with readjusted $\varepsilon_\mathrm{ss}$ and experimental data is significantly improved. This analysis suggests that if the pure component contact angles are described well, the contact angles of the mixtures can be predicted rather accurately. Future attempts to improve the presented approach should, therefore, be directed at improving the description of pure substance contact angles.

\section{Conclusion} \label{sec:conclusion}

This work provides experimental data for contact angles of various pure substance and mixtures on solid PTFE substrate and assesses predictions from a DFT model for these contact angles. Our study also includes data from \citet{li1992contact}. 
The DFT approach employs a Helmholtz energy functional based on the PC-SAFT model to describe fluid-fluid interactions. The solid-fluid interactions are included using an effective, one-dimensional external potential, where an effective energy parameter of the solid $\varepsilon_\mathrm{ss}$ is adjusted to a single contact angle measurement of $n$-octane. The approach is predictive for pure components (except $n$-octane) and for mixtures. 

We performed contact angle measurements for pure components with 43 data points for 22 unique substances, as well as for nine binary mixtures with at least six compositions for each mixture. Our study shows:
First, the DFT approach predicts rather accurately contact angles for pure substances, excluding monohydric alcohols with an average deviation to experimental values of \SI{5.1}{\degree}.
Second, a systematic overestimation of the contact angle is obtained for pure monohydric alcohols. This can be partially attributed to molecular orientation, which is not accounted for in the DFT approach. The agreement can be improved using the PC-iSAFT model, which includes the connectivity of molecular segments and, thus, accounts (to some extent) for the molecular orientation towards the solid interface.
Third, contact angles of mixtures are predicted rather accurately, whenever reasonable agreement is obtained for the respective pure substances. This includes non-polar, polar, self- and cross-associating mixtures (i.e.\ mixtures forming hydrogen bonds).  

This investigation demonstrates, that the presented DFT approach is capable of predicting contact angles of macroscopic droplets for a wide range of pure substances and mixtures. It can, therefore, be combined with continuum fluid dynamics models, which rely on contact angles as a model input and thereby provide microscopic information to macroscopic models. 

\section*{Funding}
Funded by the Deutsche Forschungsgemeinschaft (DFG, German Research Foundation) – Project Number 327154368 – SFB 1313. We thank the the German Research Foundation (DFG) for supporting this work by funding EXC 2075/1-390740016 under Germany's Excellence Strategy. We acknowledge support by the Stuttgart Center for Simulation Science (SimTech). We thank Rolf Stierle for his help improving this manuscript. We thank Prof. Dr. rer. nat. habil. Sabine Ludwigs for providing us with access to the experimental setup for the measurement of the contact angles.
\section*{Declaration of interests}
The authors have no conflict to disclose.

\section*{Data availability statement}
\textcolor{red}{The data set is currently unpublished and it will be published with the final version of this manuscript:} 
The data that support the findings of this study are openly available in the data repository of the University of Stuttgart (DaRUS) at \textcolor{red}{[Doi/Url]}, reference number \textcolor{red}{[reference number]}.

\newpage
\section*{References}

\bibliographystyle{jfm}
\bibliography{ddft_dynContactAngles}

\newpage
\section*{Supporting Information: "Static Contact Angles of Mixtures: Classical Density Functional Theory and Experimental Investigation" }

\subsection{Influence of Air on DFT Results}

\begin{figure}[ht]
  \centering
  \includegraphics[width=0.6\textwidth]{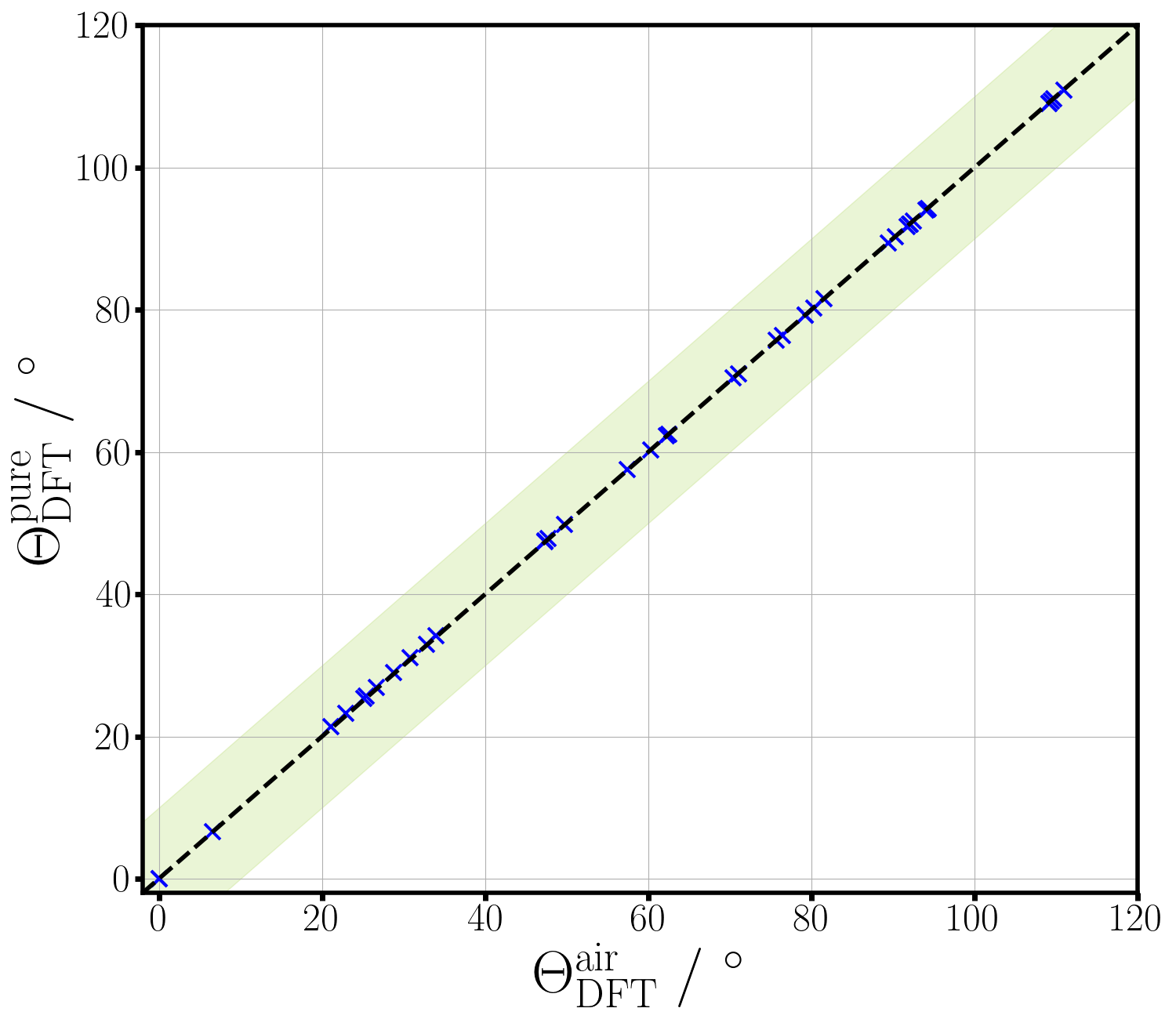} 
  \caption{Influence of the surrounding atmosphere (nitrogen/oxygen) on contact angles from DFT for pure liquids on solid PTFE substrate at $T=\SI{298}{\kelvin}$. Results with oxygen and nitrogen $\Theta_\mathrm{DFT}^\mathrm{air}$ and results assuming a pure system (neglecting the influence of oxygen and nitrogen) $\Theta_\mathrm{DFT}^\mathrm{pure}$ . 
  }
  \label{fig:pure_air}
\end{figure}

In this work, we employ a simplified approach for the calculation of contact angles from DFT by neglecting the influence of the surrounding atmosphere. For the simplified approach in the case of a pure substance, setting the temperature $T=\SI{298}{\kelvin}$ fully specifies the system. If nitrogen and oxygen are included in the DFT calculations in addition to the substance of interest, the pressure $p=\SI{1}{\bar}$ and the composition need to be specified.
\Cref{fig:pure_air} shows DFT results for contact angles of pure substances from the simplified approach ($\Theta_\mathrm{DFT}^\mathrm{pure}$, ordinate) and the more detailed approach, which explicitly includes oxygen and nitrogen in the DFT calculations ($\Theta_\mathrm{DFT}^\mathrm{air}$, abszissa). For all substances the simplified approach is in excellent agreement with results from the more detailed approach, as all  results are located close to the diagonal. This finding suggests that it is sufficient to calculate contact angles using the simplified approach.

\end{document}